\documentclass[final,twocolumn]{raa_twocolumn}           

\usepackage{graphicx,times}
\usepackage{natbib}
\usepackage{amssymb,amsmath}
\usepackage{multirow}
\usepackage{txfonts}

\bibpunct{(}{)}{;}{a}{}{,}

\usepackage[pagebackref=true,
            citecolor=blue,
            colorlinks=true,
            linkcolor=blue,
            filecolor=blue,  
            urlcolor=blue,      
            final=true,
            ]{hyperref}

\usepackage{ulem}

\usepackage{CJK}

\usepackage{xcolor}
\definecolor{addnewseccolor}{rgb}{0.5, 0.3, 0.2}

\begin{document}
\begin{CJK*}{UTF8}{gbsn}

\title{Core mass function in view of fractal and turbulent filaments and fibers
}

 \volnopage{ {\bf 2024} Vol.\ {\bf X} No. {\bf XX}, 000--000}
   \setcounter{page}{1}

\author{
Xunchuan Liu (刘训川)\thanks{liuxunchuan001@gmail.com} \inst{1,2}, Tie Liu \inst{1},
Xiaofeng Mai \inst{3,1}, Yu Cheng \inst{4},
Sihan Jiao \inst{5}, Wenyu Jiao\inst{1},
Hongli Liu \inst{6},
Siju Zhang \inst{7}
}

\institute{ 
Shanghai Astronomical Observatory, Chinese Academy of Sciences, 80 Nandan Road, Shanghai 200030, China; \\
\and
Leiden Observatory, Leiden University, P.O. Box 9513, 2300RA
Leiden, The Netherlands\\
\and 
Department of Physics, PO Box 64, 00014, University of Helsinki, Finland\\
\and
National Astronomical Observatory of Japan, 2-21-1 Osawa, Mitaka, Tokyo 181-8588, Japan\\
\and
National Astronomical Observatories, Chinese Academy of Sciences, 20A Datun Road, Chaoyang District, Beijing 100012, China\\
\and
School of physics and astronomy, Yunnan University, Kunming, 650091, China\\
\and
Departamento de Astronomía, Universidad de Chile, Las Condes, 7591245
Santiago, Chile\\
\vs \no
   {\small Received 20XX Month Day; accepted 2025 Jan 29}
}

\abstract{
We propose that the core mass function (CMF) can be driven by filament fragmentation. To model a star-forming system of filaments and fibers, we develop a fractal and turbulent tree with a fractal dimension of 2 and a Larson's law exponent (\(\beta\)) of 0.5.
The fragmentation driven by convergent flows along the splines of the fractal tree yields a Kroupa-IMF-like CMF that can be divided into three power-law segments with exponents $\alpha$ = $-0.5$, $-1.5$, and $-2$, respectively. The turnover masses of the derived CMF are approximately four times those of the Kroupa IMF, corresponding to a star formation efficiency of 0.25. Adopting $\beta=1/3$, which leads to  
fractional Brownian motion along the filament, may explain 
a steeper CMF at the high-mass end, with $\alpha=-3.33$ close to 
that of the Salpeter IMF. We suggest that the fibers of the tree are basic building blocks of star formation, with similar properties across different clouds, establishing a common density threshold for star formation and leading to a universal CMF.
\keywords{stars: formation --- stars: kinematics and dynamics --- turbulence --- stars: mass function --- ISM: clouds
}
}

   \authorrunning{Liu, et al.}            
   \titlerunning{CMF via filament fragmentation}  
   \maketitle

%
\section{Introduction}           
\label{sect:intro}
The Core Mass Function (CMF) is a fundamental concept in the study of star formation. Observations indicate that the CMF typically follows a power-law distribution \citep[e.g.,][]{1998A&A...336..150M,2015A&A...584A..91K,2018ApJ...853..160C,2022A&A...664A..26P}, suggesting a potential link between the CMF and the Initial Mass Function \citep[IMF,][]{1955ApJ...121..161S,2001MNRAS.322..231K,2007A&A...462L..17A,2015MNRAS.450.4137G}.
Exploring the origin of CMF is a key to
finding out the fundamental physics that regulate 
IMF and star formation \citep{2022A&A...662A...8M}.
Various theories have been proposed to construct the CMF. Some of the most influential models \citep[e.g.,][]{2008ApJ...684..395H,2012MNRAS.423.2037H,2018ApJ...854...35H} begin with the fragmentation of density fluctuations that follow a lognormal distribution, which are produced by three-dimensional supersonic shocks \citep[e.g.,][]{2002ApJ...576..870P}.
These theories focus on general three-dimensional isotropic cases and do not account for the filamentary nature of core-forming clouds or clumps, such as their anisotropy, self-similarity, and low dimensionality.

Filaments are believed
to play a very important role in star formation \citep[e.g.,][for recent reviews]{2023ASPC..534..153H,2023ASPC..534..233P}.
Filaments are elongated structures (typically longer than 1 pc) that can extend through molecular clouds/clumps.
{\it Herschel} studies of nearby low-mass star-forming regions 
\citep[e.g.,][]{2010A&A...518L.102A,2010PASP..122..314M,2012A&A...541A..12J}
show that 
filaments dominate the mass budget of molecular clouds 
and correspond to the birthplaces of most prestellar cores 
\citep[e.g.,][]{2010A&A...518L.102A,2014ApJ...791...27S,2015A&A...584A..91K}.
Velocity gradients are common along the fialments.
The interferometers make it possible to resolve both nearby and distant filament systems down to core scales \citep[e.g.,][]{2018A&A...617A.100B,2020MNRAS.496.2790L,2022A&A...662A...8M,2024RAA....24b5009L,2024A&A...690A.185W}.
The distant, massive, dense clumps are also found to consist of hierarchical filamentary hubs and networks \citep[e.g.,][]{2013A&A...555A.112P,2022MNRAS.514.6038Z,2023ApJ...953...40Y}.
The filament systems can be further divided into fibers in both low- \citep{2018A&A...610A..77H} and intermediate- \citep{2024ApJ...976..241Y}, as well as high-mass \citep{2024RAA....24b5009L,2024A&A...687A.140H} star-forming regions.
Fibers are small branches (typically shorter than 1 pc) of filaments that can only be effectively resolved by interferometers
\citep[e.g.,][]{2024A&A...687A.140H}.
One of the most important characteristics of fibers is that they may exist in subsonic states \citep{2021RAA....21...24Y}.

Theories of CMF that emphasize the unique role of filaments/fibers are crucial for unveiling the 
underlying nature of star formation. 
\cite{2013ApJ...764..140M} explained the CMF through radial accretion onto filaments by introducing an exponential-like stopping time in the accretion process. They presumed that the CMF has a shape similar to that of the IMF. Another limitation of the model is that \cite{2013ApJ...764..140M} have not considered the fragmentation or the turbulent structure along the filaments, making the model less specific to filamentary structures.
\citet{2015A&A...584A.111R} linked the power-law spectrum of density fluctuations along the filament to the core mass function. However, they did not explain the origin of the power spectrum in detail, nor how it may be influenced by different turbulence structures. Moreover, they did not consider the fractal nature of filaments. 
Overall, it is important to establish a theoretical model that can produce an IMF-like CMF while fully accounting for the fractal and turbulent characteristics of filament networks. 


In this work, we propose a theoretic model of filament fragmentation that successfully constructs an IMF-like CMF. The paper is organized as follows: In Sect. \ref{sec_whycommom}, we explain why it is theoretically possible to derive the CMF by studying filaments. In Sect. \ref{sec_cmfforfil}, we develop a model of fractal and turbulent trees to analogize filament/fiber systems, and construct a Kroupa-IMF-like CMF through filament fragmentation. Further,  we discuss how to construct a steeper (Salpeter-IMF-like) CMF at the high-mass end based on fractional Brownian motion (Sect. \ref{sec_fbm_hcmf}), explain the universality of the CMF (Sect. \ref{sec_why_univ}),
and present the assumptions and caveats of the model proposed in this work (Sect. \ref{sec_assum_caveats}). In Sect. \ref{sec_summary}, we provide a brief summary.

\section{Why are filaments a stable texture}\label{sec_whycommom}
The dimension ($n = 0, 1, 2$) of a core, filament, or sheet is defined by the dimensionality of its central part: point, line, or plane.
The typical radius ($a$) and central density ($\rho_0$) of an isothermal $n$-dimensional ($n$-D) structure are related by the following equation (Eq.~\ref{eq_comwidth}): 
\begin{equation}
    a \propto \rho_0^{-1/2}, \label{eq_comwidth_maintxt}
\end{equation}
where the power-law index is independent of the dimension.
The mass (or line mass/surface density) enclosed within a radius of $a$ for the \( n \)-D structures 
(\( n = 0, 1, 2 \)), denoted as \( \eta_{nd} \), can be estimated as:
\begin{equation}
    \eta_{nd} \sim \rho_0 a^{3 - n} \propto \rho_0^{(n - 1)/2}. \label{eq_rhon}
\end{equation}
For $n=2$, it is evident that \( \eta_{2d} \) is an increasing function of \( \rho_c \).
This implies that 
reducing the scale \( a \) (and thus increasing \( \rho_0 \)) of 
a sheet
will require a higher surface density to make it gravitationally bound.
Therefore, a 2D sheet is difficult to compress continuously under its own self-gravity. 
For $n = 0$, the solution to Eq.~\ref{eq_profi} is the well-known Bonnor-Ebert (B-E) sphere \citep{1956MNRAS.116..351B}, for which $\eta_{0d} \propto \rho_0^{-1/2}$.
When a sphere in equilibrium is slightly compressed due to perturbations, the critical mass becomes smaller than the real mass, triggering an accelerating collapse.

For $n = 1$, which corresponds to the case of an isolated filament, \( \eta_{1d} \) is independent of \( r \) (Eq. \ref{eq_rhon}). 
In equilibrium, the radial density profile is described by the Plummer function \citep{1911MNRAS..71..460P}, as confirmed by observations \citep[e.g.,][]{2008MNRAS.384..755N,2011A&A...529L...6A,2021ApJ...912..148L,2022A&A...667L...1A}.
The precise value of the critical line mass of a filament is given by 
\citep{1964ApJ...140.1056O,1997ApJ...480..681I,2014prpl.conf...27A}
\begin{equation}\label{eq_clm}
    \eta_{1d}^{\rm crit} = \frac{2 \delta^2}{G}.
\end{equation}
Here, in the case of thermal support, \( \delta \) represents the sound speed \( (c_s) \); 
for turbulent support, the turbulent velocity dispersion ($\delta_{\rm turb}$) should
be taken into account: $\delta^2=c_s^2+\delta^2_{\rm turb}$. 
As an example calculation of \( \eta_{1d}^{\rm crit} \), for \( \delta = 0.3 \, \text{km s}^{-1} \),  \( \eta_{1d}^{\rm crit} \sim 40 \, M_\odot \, {\rm pc}^{-1} \).

Overall, two-dimensional sheets are difficult to be compressed by self-gravity.
One-dimensional filaments and zero-dimensional cores are two key types of self-gravitating structures. The instability of cores finally leads 
to formation of stars. However, a filament cannot undergo a sustained global collapse. 
Observations show that filaments can be disrupted by strong impacts, such as the expansion of H{\sc ii} regions \citep[e.g.,][]{2016A&A...585A.117Z,2022MNRAS.514.6038Z}. However, even under such strong impact, the clumps/cores in the compressed gas may still originate from the sweeping up of pre-existing fragments \citep[e.g.,][]{2024MNRAS.tmp.2359Z}.
These analyses, supported by observational evidence, suggest that filaments are stable structures, resistant to disruption by both internal gravity and external pressures (e.g., turbulence and stellar feedback). 
As a result, they help preserve the initial seeds of star formation.

\section{CMF of filaments}\label{sec_cmfforfil}
\subsection{Fragmentation along a filament of infinite length}
\label{sec_inffil}
Larson's law \citep{1981MNRAS.194..809L} yields that
\begin{equation}
    \sigma \propto L^\beta \label{eq_turbst}
\end{equation}
where 
$\sigma$ is the velocity dispersion and 
$L$ is the size of the cloud.
The original value of \( \beta \) given by \cite{1981MNRAS.194..809L} is 0.38, which is close to the value (1/3) expected for standard Kolmogorov turbulence \citep{1941DoSSR..30..301K,1995tlan.book.....F}.
For compressible fluid (Burgers turbulence), 
$\beta=0.5$ \citep{BURGERS1948171}. We adopt \(\beta = 0.5\) for the turbulent and gravity-bound star-forming clouds \citep[e.g.,][]{2009ApJ...699.1092H}.

\begin{figure*}
    \centering
    \includegraphics[width=0.78\linewidth]{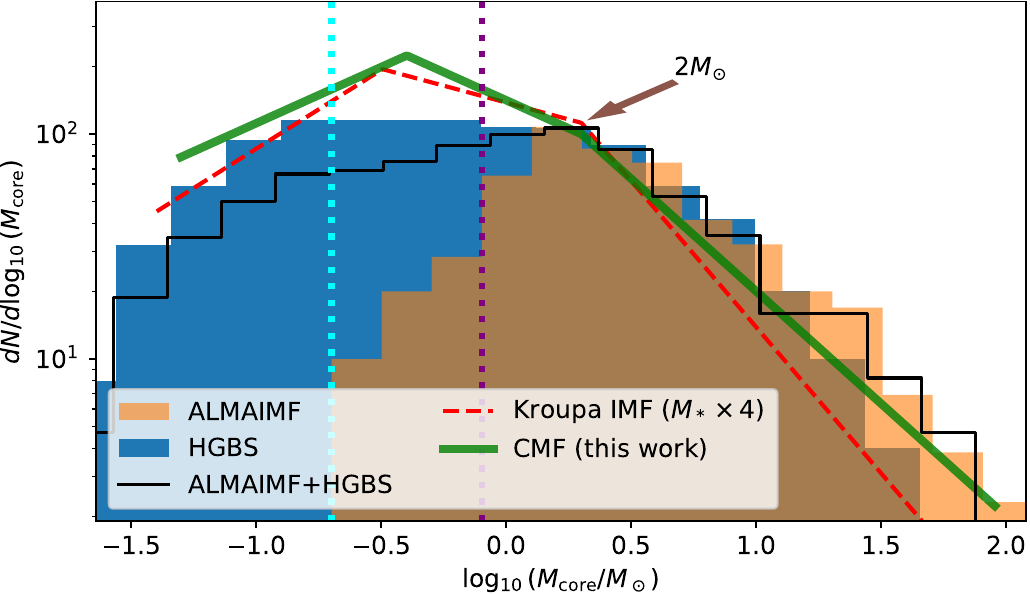}
    \hspace{0.03\linewidth}
    \caption{The core mass function (CMF) of the ALMA-IMF survey \citep{2022A&A...662A...8M,2024A&A...690A..33L} is shown in orange, and that of the {\it Herschel} HGBS survey \citep{2010A&A...518L.102A,2015A&A...584A..91K} in blue. We adopt the full core catalog of the ALMA-IMF survey for all 15 massive star formation regions, and the core catalog for the Aquila cloud complex covered by HGBS. 
    The cyan and purple dotted vertical lines indicate the 90-percent-completeness core mass for the HGBS and ALMA-IMF surveys, respectively.
    The CMFs of both datasets exhibit a similar turnover mass at $\sim 2$ $M_{\sun}$, consistent with the value predicted in this work (Sect. \ref{sec_mergeends}). The CMFs of the two datasets also coincidentally have a similar number of cores around the turnover mass. Therefore, we combine them directly with equal weightings. The black line represents the CMF of the combined dataset (with the y-axis values adjusted so that it reaches 100 at $2\, M_{\sun}$). The red dashed line represents the shifted Kroupa IMF, with the stellar mass multiplied by a factor of four. The values of \( \alpha \) for the shifted Kroupa IMF within the three intervals (from left to right) are $-0.3$, $-1.3$, and $-2.3$, respectively \citep{2001MNRAS.322..231K}.
     The green line represents the theoretical CMF of this work (Eq. \ref{eq_cmf_our}).
    \label{fig_cmf}
    }
\end{figure*}

An isolated filament can be treated as a 1-D structure, with the velocity along the filament following the turbulent motion described by Larson's law (Eq. \ref{eq_turbst}).
The velocity distribution along the filament can then be modeled as 1-D random walk in Brownian motion, 
\begin{equation}
    \langle v^2 \rangle \propto L.
\end{equation}
The probability distribution of the zero-crossing intervals \( L_{\text{Z}} \) follows \citep{wiener1923,gardiner2009}
\begin{equation}
P(L_{\text{Z}}) \propto L_{\text{Z}}^{-1.5}, \label{eq_BM_PL}
\end{equation}
and the average number of the zero-crossing events (\( \langle N_{\text{Z}} \rangle \)) within an interval of length \( L \) is given by
\begin{equation}
\langle N_{\text{Z}}(L) \rangle \propto L^{0.5}.\label{eq_Ncross}
\end{equation}
We assume that a filament is radially supported by thermal dispersion
with a nearly constant $\eta_{1d}^{\rm crit}$. 
The fragmentation of the filament is driven by convergent flow 
along the filament around the zero-crossing points of the
tangential velocity. 
The mass of the fragmentation should be proportional to $L_Z$. 
The mass function of the fragmentation along a filament of infinite length (FMF) is (Eq. \ref{eq_BM_PL})
\begin{equation}
    {\rm FMF} (M) \propto M^{-1.5}. \label{eq_FMF}
\end{equation}
The FMF is noticeably flatter than the observed CMF and IMF at the high-mass end \citep[e.g.,][see also Fig. \ref{fig_cmf}]{2019A&A...629L...4A}. Moreover, the integral of the above equation diverges. In realistic scenarios, there should be cutoffs at both the high-mass and low-mass ends, denoted as \( M_{\rm min} \) and \( M_{\rm max} \), respectively. We adopt:
\begin{align}
    M_{\rm min} \sim \eta_{1d}^{\rm crit} \mathcal{W},\\
    M_{\rm max} \propto \eta_{1d}^{\rm crit} L.\label{eq_mmax}
\end{align}
where $\mathcal{W}$ is the width of the filament.

\subsection{CMF of a fractal tree of filaments}\label{sec_maxcore}
\subsubsection{A tree embedded in a star-forming cloud/clump}
For a massive clump supported by turbulence as described by Eq. \ref{eq_turbst}, the viral equilibrium requires
\begin{equation}
\frac{GM_C}{r_C}  \propto r_C,
\end{equation}
where \( M_C \) and \( r_C \) represent the mass and radius of the clump, respectively. It yields
\begin{equation}
    r_C \propto M_C^{{1}/{2}} \label{eq_rc}.
\end{equation}
The density profile of the clump ($\rho_C$) is 
\begin{equation}
\rho_C \sim M_C/r_C^3 \propto r_C^{-1}.\label{eq_clrho}
\end{equation}
We assume that a star-forming cloud/clump is dominated by filamentary
structures composed of fibers that are thermally supported, with similar critical line masses.
It implies that the filament/fiber system should have a fractal dimension (\( D_{\text{fil}} \)) given by
\begin{equation}
    D_{\text{fil}} = \frac{\log(M_C)}{\log(r)}  = 2.\label{eq_Dfil}
\end{equation}

\subsubsection{Construct the fractal tree}
\label{sec_lowmassend}\label{sec_frac_fil}
The fractal structure of the filament/fiber system plays a critical role in regulating the statistical properties of core masses, and thus the FMF in Eq. \ref{eq_FMF} cannot be directly used to construct the CMF. 
Here, we explore the CMF 
through constructing a model of 
a self-similar tree.
The basic blocks of a tree are fibers with constant length ($\mathcal{L}$) and line mass. These fibers merge progressively to form increasingly larger branches (sub-trees), ultimately shaping the tree.
The widths of the fibers ($\mathcal{W}$) may decrease as they evolve.
The fibers along which cores/stars begin to appear are denoted as \( \mathcal{T}_1 \) and its length is denoted as $\mathcal{L}_1$. 
Thus, the entire tree represents a core/star-forming structure in a dynamically evolving process.
Each sub-stree has a longest path, denoted as the spline of the sub-tree. A sub-tree is denoted as $\mathcal{T}_i$ if its spline has a length of $\mathcal{L}_i=i\times \mathcal{L}_1$. Here, $i$ is a natural number ($i = 1, 2, 3, \dots$).
The whole tree, with a spline length of
$\mathcal{L}_{\mathcal{N}}=\mathcal{N}\times \mathcal{L}_1$, is denoted as $\mathcal{T}_{\mathcal{N}}$.
We require 
\begin{equation}
    \mathcal{L}_{\mathcal{N}} \sim R_C,\label{eq_LN}
\end{equation}
where $R_C$ is the cloud/clump size.
Eqs. \ref{eq_mmax} and \ref{eq_LN}, as well a constant $\eta_{1d}$, imply that the maximum core mass of the clump (\( M_{\rm max}^C \)) follows
\begin{equation}
    M_{\rm max}^C \propto M_C^{1/2}. \label{eq_mmaxc}
\end{equation}
This is consistent with the results of simulations  \citep[e.g.,][]{2004MNRAS.349..735B} and observations 
\citep[e.g.,][]{2021MNRAS.508.2964A,2023ApJ...950..148M,2023A&A...670A.151Y}.
A recent statistical result based on data from the {\it Herschel} 
HGBS survey
\citep{2010A&A...518L.102A} and the ALMA-IMF survey \citep{2022A&A...662A...8M} also yields a power-law index $\sim$0.5
(Jiao et al., submitted; private communication).

To construct the tree, we proceed from $\mathcal{T}_1$
at the tail of the spline of $\mathcal{T}_{\mathcal{N}}$. 
At the $i_{\rm th}$ step, the head of $\mathcal{T}_i$ would intersect with the heads of smaller sub-trees ($\mathcal{T}_j$ with $j<i\leq \mathcal{N}$). 
Each \( \mathcal{T}_j \) appears with a probability given by
\begin{equation}
    p_j = \frac{b j^{-\gamma}}{\zeta(\gamma)}.
\end{equation}
Here, \( \zeta(\gamma) = \sum_j j^{-\gamma} \), and \( b \) can be interpreted as the average number of sub-trees that join at each step.
Denote $\mathcal{M}_i$ as the mass of $\mathcal{T}_i$. We have
\begin{equation}
    \mathcal{M}_{i+1}  \sim \mathcal{M}_i + 1 + 
    \sum_1^{i} p_j \mathcal{M}_j.
\end{equation}
$\mathcal{M}_i$ should be proportional to \( \mathcal{L}_i^{D_{\rm fil}} \) (Eq. \ref{eq_Dfil}), that is
\begin{equation}
    \mathcal{M}_i  
    \propto \mathcal{L}_i^{2}, \label{eq_Mii}
\end{equation}
which is consistent with  the statistical results of the observations \citep{2023ASPC..534..153H}.
It leads to
\begin{equation}
   (i+1)^{D_{\rm fil}} \propto i^{D_{\rm fil}}+ i^{D_{\rm fil}+1-\gamma}b/\zeta(\gamma)/(D_{\rm fil}+1-\gamma),
\end{equation}
where $1<D_{\rm fil}\leq 2$.
It requires that $\gamma=2$, and $D_{\rm fil}=b/\zeta(2)/(D_{\rm fil}-1)$, which leads to
\begin{equation}
  D_{\rm fil}=  \sqrt{b/\zeta(2)+0.25}+0.5    
\end{equation}
Adopting \( b = 2\zeta(2) \) yields a \( D_{\rm fil} \) of 2.

\subsubsection{CMF at the high-mass end}\label{sec_cmf_he}
On average, the entire tree \( \mathcal{T}_{\mathcal{N}} \) can be divided into \( N_i \) sub-trees \( \mathcal{T}_i \), where \( N_i \) is estimated from Eq. \ref{eq_Mii} as:
\begin{equation}
  N_i \sim \left( \frac{\mathcal{N}}{i} \right)^{D_{\rm fil}}
  \propto \mathcal{L}_i^{-D_{\rm fil}}.
  \label{eq_filwid}
\end{equation}
It directly yields a spline length function (SLF) of
\begin{equation}
    {\rm SLF}(\mathcal{L}) \propto \mathcal{L}^{-2}.\label{eq_SLF}
\end{equation}
The high-mass cutoff for the CMF 
along the spline and across the entire branches of $\mathcal{T}_i$,
derived from Eq. \ref{eq_mmax}, is
\begin{equation}
     M_{\rm max}^{i} \propto \mathcal{L}_i \propto i. \label{eq_Mmaxi}
\end{equation}
Eqs. \ref{eq_Ncross}, \ref{eq_FMF}, \ref{eq_SLF}, and \ref{eq_Mmaxi} lead to a high-mass-end CMF (denoted as ${\rm CMF}^{\rm H}$):
\begin{align}
    {\rm CMF}^{\rm H}(M) &\propto \int_{{M}}
    x^{-2}  x^{0.5} 
    {\rm FMF}(x) dx \nonumber \\
    &\propto M^{-2}. \label{eq_CMF_h}
\end{align}
It means that the power-law index of the CMF at the high-mass end
is determined by that of the SLF. The massive cores are likely to be situated at the convergent hubs, where different branches intersect.
    
\subsubsection{CMF at the low-mass end}
We assume that the density along the spline of a $\mathcal{T}_i$ (denoted as $\rho_i$) follows 
\begin{equation}
    \rho_i \propto i^p, 
\end{equation}
and  $\rho_\mathcal{N}$ and  $\rho_1$ are proportional to
the densities at the central and boundary of the cloud/clump, respectively.
From Eq. \ref{eq_clrho}, we have 
\begin{equation}
    \rho_\mathcal{N}/\rho_1 \propto \mathcal{N},
\end{equation}
with further requires that 
\begin{equation}
    \rho_i \propto i. \label{eq_rhoi}
\end{equation}
From Eq. \ref{eq_comwidth_maintxt}, the width  of the spline of $\mathcal{T}_i$ (denoted as $\mathcal{W}_i$) can be expressed as:
\begin{equation}
    \mathcal{W}_i \propto i^{-0.5}.\label{eq_widthi}
\end{equation}
It sets a low-mass cutoff for the CMF along the spline and across the entire branches of $\mathcal{T}_i$ as:
\begin{equation}
    M_{\rm min}^{i} \propto \mathcal{W}_i \propto i^{-0.5}
    \label{eq_core_lowmasscut}
\end{equation}
Note that $M_{\rm min}^{i}$ can be interpreted as the thermal Jeans mass corresponding to $\rho_i$.
Low-mass cores with masses \( \sim M_{\rm min}^{i} \) are primarily produced along the spline of \( \mathcal{T}_i \).
The CMF at the low-mass end (CMF$^{L}$) can then be derived from Eqs. \ref{eq_Ncross}, \ref{eq_Dfil}, \ref{eq_filwid}, \ref{eq_widthi}, and \ref{eq_core_lowmasscut} as:
\begin{equation}
    {\rm CMF^L}(M) \propto  N_i\left(\frac{i}{\mathcal{W}_i}\right)^{0.5} \left(\frac{dM_{\rm min}^{i}}{di}\right)^{-1}
    \propto  M^{-0.5}. \label{eq_smallendcmf}
\end{equation}
The longest (and thus most evolved) spline of the tree contributes to the formation of cores with both the lowest and highest masses. 
The power-law index of the derived ${\rm CMF}^{\rm L}$ is
compatible with that of the Kroupa IMF at the low-mass end ($-0.3$ to $-1.3$).

\subsection{The merged CMF and its turnover masses} \label{sec_mergeends}

It is not straightforward to merge CMF$^{\rm L}$ and CMF$^{\rm H}$, as most of the quantities discussed are in very general cases without characteristic scales. This is somewhat reasonable, as both filaments and fragmentation are universal phenomena that span a wide range of scales. For filament/fiber systems within star-forming clouds/clumps that we focus on here, there exist some characteristic scales. 
The fractal tree constructed above has two important characteristic 
scales: the typical width and length of $\mathcal{T}_1$,
which determine the $M_{\rm min}^1$ and  $M_{\rm max}^1$,
respectively.
The CMF is expected to be divided into three segments by $M_{\rm min}^1$ and $M_{\rm max}^1$, with each segment having a different power-law index ($\alpha$).
The power-law indices at the low-, intermediate-, and high-mass ends are adopted as $-0.5$ (Eq. \ref{eq_smallendcmf}), 
$-1.5$ (Eq. \ref{eq_FMF}), and $-2$ (Eq. \ref{eq_CMF_h}), respectively.

The filament length and width of $\mathcal{T}_i$ are uncertain.
We assume that the `typical width (0.1 pc)' of
the {\it Herschel} filaments \citep{2022A&A...667L...1A} is contributed to the elemental fibers ($\mathcal{T}_1$)
that intersect with the splines of the filaments, with a
length-to-width aspect ratio of five.
Then, $\mathcal{L}_1$ and
$\mathcal{W}_1$  are estimated to be 
10\,000 au ($\sim 0.05$ pc) and 2\,000 au, respectively.
This is consistent with the observational results from SMA and ALMA for nearby clouds and distant massive clumps \citep[e.g.,][]{2021RAA....21...24Y,2024ApJ...976..241Y}.
A typical temperature of 25 K of molecular gas 
corresponds to a critical line mass $\eta_{1d}^{\rm crit}=40$ 
$M_{\sun}$ pc$^{-1}$ \citep[e.g.,][]{2024ApJ...976..241Y}.
We thus further estimate 
\begin{equation}
 \left\{
    \begin{aligned}
   &M_{\rm max}^1 \sim \eta_{1d}^{\rm crit} \mathcal{L}_1 \sim 2\,M_{\sun},\\
   &M_{\rm min}^1 \sim \eta_{1d}^{\rm crit} \mathcal{W}_1 \sim 0.4\,M_{\sun}.   
    \end{aligned}
    \right. \label{eq_M1cuts}
\end{equation}
Then we have
\begin{equation}
 {\rm CMF}(M) =\left\{
    \begin{aligned}
    &M^{-0.5} && {\rm for}\,M<0.4\,M_{\sun} \\
    &M^{-1.5} && {\rm for}\,0.4<M<2\,M_{\sun} \\
    &M^{-2}  && {\rm for}\,M>2\,M_{\sun} 
    \end{aligned}
    \right. \label{eq_cmf_our}
\end{equation}
The turnover mass $M_{\rm max}^1\sim 2\, M_{\sun}$ is
consistent with the CMFs given by the observations of the
{\it Herschel} HGBS survey \citep{2022A&A...662A...8M,2024A&A...690A..33L} and the ALMA-IMF survey \citep{2010A&A...518L.102A,2015A&A...584A..91K} (Fig. \ref{fig_cmf}).
The CMF in Eq. \ref{eq_cmf_our} is also similar to that of the Kroupa IMF \citep{2001MNRAS.322..231K}, except for a multiplication of the star mass of the Kroupa IMF by a factor of four, which corresponds to a core-to-star mass efficiency of 25 percent (Fig. \ref{fig_cmf}).

The small discrepancy between the derived CMF (Eq. \ref{eq_cmf_our})
and the shifted  Kroupa IMF may contribute to the non-linear relationship between the CMF and the IMF
\citep[e.g,][]{2013MNRAS.432.3534H,2015MNRAS.450.4137G}.
The overpopulation of the derived CMF, compared to the observations (Fig. \ref{fig_cmf}), may contribute to the miscounting of the flattened cores of $\mathcal{T}_1$, due to the limited angular resolution of single-dish telescopes and the missing flux effect of interferometers \citep[e.g.,][]{2023ApJ...945..156S}.


\section{Discussion}
\subsection{Fragmentation under fractional Brownian motion}
\label{sec_fbm_hcmf}
Above, CMF$^{\rm H}$ is obtained by convolving the FMF and SLF, with the result depending solely on the SLF (Sect. \ref{sec_cmf_he}).
It matches observations of massive star-forming regions, but is slightly flatter than that of nearby low-mass star-forming regions (Fig. \ref{fig_cmf}). Here, we note that steeper CMF$^{\rm H}$ can be reproduced by the turbulence fragmentation of isolated filaments.

In Sect. \ref{sec_inffil}, we considered only $\beta = 0.5$. For other values of $\beta$ (e.g., $\beta = 0.33$ for Kolmogorov turbulence), we modeled the velocity distribution along the filament as fractional Brownian motion (fBM), 
\begin{equation}
    \langle v^2 \rangle \propto L^{2\mathcal{H}},
\end{equation}
where \( \mathcal{H} = \beta \) is the Hurst exponent.
Similar to Eqs.~\ref{eq_BM_PL} and \ref{eq_Ncross}, for fBM \citep{mandelbrot1968,RAMBALDI199421,biagini2008}, the probability distribution of the zero-crossing intervals ($P_\mathcal{H}(L_{\text{Z}})$) and the average number of zero-crossing events ($\langle N_{{\text{Z};\mathcal{H}}}(L) \rangle$) satisfy:
\begin{equation}
P_\mathcal{H}(L_{\text{Z}}) \propto L_{\text{Z}}^{-(2-\mathcal{H})}, \label{eq_fBm_L}
\end{equation}
\begin{equation}
\langle N_{{\text{Z};\mathcal{H}}}(L) \rangle \propto L^{1-\mathcal{H}} \label{eq_fBm_frag},
\end{equation}
respectively.
The first-level fragmentation by converging flows along an isolated
filament (Eq. \ref{eq_fBm_L}) leads to a power-law index of FMF
of $-(2\!-\!\mathcal{H})$.
The local velocity gradient could be steeper than the
global velocity gradient around the zero-crossing points,
leading to second-level fragmentation. 
We assume that second-level fragmentation tends to break a segment of length \(L\) into one larger sub-segment, with a length distribution following Eq. \ref{eq_fBm_L}, along with several smaller sub-segments.
At the long-segment end, the length of the sub-segment (\(l\)) follows a probability distribution given by
\begin{equation}
    {\rm FMF_{2}}(l) \propto \frac{P_\mathcal{H}(l)}{\langle N_{{\text{Z};\mathcal{H}}}(l) \rangle} \propto l^{-(3-2\mathcal{H})},
\end{equation}
which results in a CMF$^{\rm H}$ with an exponent \(\alpha = -(3\!-\!2\mathcal{H})\).
For Kolmogorov turbulence ($\mathcal{H}=\beta=1/3$), $\alpha=-2.33$, very
close to the power-law index of the Salpeter IMF ($-2.35$)
\citep{1955ApJ...121..161S} and the Kroupa IMF ($-2.3$) at the
high-mass end. 
For Burgers turbulence ($\mathcal{H}=\beta=1/2$),
it again gives a top-heavy CMF with $\alpha=-2$.
These results suggest that the CMF at the high-mass end may be linked to some fundamental and universal mechanisms of star formation. 
We propose an \(\alpha\) between \(-2\) and \(-(3 - 2\beta)\), depending on how the structures of the turbulence and the filament system are formed.

\subsection{Why the fractal tree is universal}\label{sec_why_univ}
The column density of the splines of $\mathcal{T}_i$ (denoted as $\Sigma_i$) is (Eqs. \ref{eq_rhoi} and \ref{eq_widthi})
\begin{equation}
    \Sigma_i \propto \rho_i\mathcal{W}_i\propto i^{0.5}.
\end{equation}
The distribution of $\Sigma_i$ can be expressed as
\begin{equation}
    dN(\Sigma_i)  \propto N_i\mathcal{L}_i\mathcal{W}_i di \propto \Sigma_i^{-1} d\log(\Sigma_i),
\end{equation}
yielding a column density
probability distribution function ($N$-PDF) power-law index ($\omega$) of $-1$.
Observations show that the value of $\omega$, at the high-column-density end influenced by star formation \citep[e.g.,][]{1998ApJ...504..835S,2009A&A...508L..35K}, exhibits a decreasing trend  (from $-1$ to $-3$) over the evolution of the cloud/clump, reaching a value of $-1$ in the early stages \citep[e.g.,][]{2015A&A...577L...6S}.
It confirms that the fractal tree we have established captures the key properties of the early-stage structures that harbor the seeds of star formation.

The universality of the turn-over masses of the CMF derived in Sect. \ref{sec_cmfforfil} relies on that of \(\mathcal{T}_1\) (Eq. \ref{eq_M1cuts}).
The host cloud/clump of $\mathcal{T}_i$
has an average volume density ($n_i$) of 
\begin{equation}
    n_{\rm Ci} \sim M_i/\mathcal{L}_i^3 \sim n_{\rm C1}/i.
\end{equation}
with
\begin{equation}
    n_{\rm C1} \sim \eta_{1d}^{\rm crit}/\mathcal{L}_1^2 \sim 10^5\ {\rm cm^{-3}}.
\end{equation}
The mass of a star-forming tree ($\mathcal{M}_{\mathcal{N}}$) is dominated by the mass of all the
$\mathcal{T}_1$ branches of the tree. The star formation rate (SFR) follows 
\begin{equation}
    {\rm SFR} \propto \mathcal{M}_{\mathcal{N}} \propto M_{n>n_{\rm C1}},
\end{equation}
where
\begin{equation}
   M_{n>n_{\rm C1}} = \int_{n>n_{\rm C1}} \rho dV. 
\end{equation}
We assume that $n_{\rm C1}$ is the density threshold above which star-forming trees can be established, starting from $\mathcal{T}_{1}$.
If these trees share similar density thresholds and evolutionary processes, a universal density threshold for star formation may be expected.
This is consistent with the concept of the
well-known star formation law, which describes the linear correlation between dense gas surface density and star formation rate \citep{1998ApJ...498..541K,2004ApJ...606..271G}. Commonly used dense gas tracers, such as HCO$^+$ (1-0) and HCN (1-0) \citep[e.g.,][]{2004ApJ...606..271G,2017A&A...604A..74S,2022A&A...667A..70R}, have critical densities \citep[approximately $10^5$ cm$^{-3}$;][]{2015PASP..127..266M} similar to the value of $n_{\rm C1}$.
We thus propose that the universality of the CMF is linked to that of the star formation law.

\subsection{Assumptions and caveats}\label{sec_assum_caveats}
Here, we list the key assumptions made in establishing the model of this work.
(1) The density and turbulent structures of a cloud/clump are dominated by those of a hierarchically structured filamentary tree. (2) The tree can be resolved into fibers that are radially supported by thermal motion. (3) Although the critical line masses of fibers and splines are maintained throughout the evolution of the tree, their density gradually increases. (4) Jeans fragmentation of the evolved splines and convergent flows along them contribute to the formation of low-mass and high-mass cores, respectively.

The first assumption posits that the 1-D fibers merge into 2-D fractal filamentary trees that fully extend within the clouds/clumps.
Since a 2-D sheet is difficult to compress under self-gravity (Sect. \ref{sec_whycommom}), the filamentary trees play the roles of both 2-D and 1-D structures during the process of gas collection from 3-D clouds/clumps into point-like protostars.
The second assumption is pretty strong.
Fibers in low-mass clouds have been observationally shown to be subsonic \citep{2021RAA....21...24Y}, which is a key characteristic of fibers. Filaments in intermediate-mass star-forming regions can also be resolved into fibers that are narrow in line width \citep{2024ApJ...976..241Y}. Some observational evidence suggests that filaments in star-forming regions with high-mass young protostars could also be resolved into subsonic fibers by ALMA \citep{2024A&A...687A.140H}. However, to date, no systematic observational studies of remote high-mass star-formation regions have been conducted on this issue, and we remain to treat it as an assumption.
The third assumption depends not only on the spatial distribution of the fibers, but also on how they dynamically merge into a filamentary tree.
The detailed processes remain poorly understood, both observationally and theoretically.
The fourth assumption is essentially the key concept of this model.
One of the inferences of the fourth assumption is that a young massive protostar should be accompanied by some low-mass protostars of similar evolutionary age around its birthplace. This is supported by the recent detection of both a low-mass hot corino and a high-mass hot core within the same protocluster \citep{2023ApJ...958..174L}.

There are several caveats associated with the model presented in this work. Notably, the potentially important role of magnetic fields has not been considered. Molecular clouds with strong magnetic fields may require a higher threshold density for the formation of star-forming filamentary structures, thereby further altering the CMF.
For simplicity, we have also not considered the effects of feedback from high-mass protostars. Although filaments can be resistant to disruption by both internal gravity and external pressures (Sect. \ref{sec_whycommom}), the strong influence of expanding H{\sc ii} regions may alter the topology and core mass function (CMF) of the filaments. Additionally, the thermal feedback from protoclusters may lead to higher core masses by increasing the critical line masses of filaments and fibers. 
At present, this model remains in the realm of theoretical exploration and requires further observations to help validate its assumptions and inferences.

\section{Summary}\label{sec_summary}
In this work, we propose that the filament system is a stable structure that preserves the initial seeds of star formation. We model a fractal and turbulent tree of filaments/fibers with a fractal dimension of 2 and a Larson's law exponent (\(\beta\)) of 0.5 to explore the CMF under turbulent fragmentation along the filaments. A Kroupa-IMF-like CMF can be produced. The CMF at the low-mass end (\(<0.4 \, M_{\sun}\)) is influenced by the turbulent fragmentation of sub-trees at different evolutionary stages, and thus by variations in filament widths. The CMF at the high-mass end (\(>2 \, M_{\sun}\)) depends on the fractal dimension of the tree. By adopting a \(\beta\) of 1/3, it is possible to construct a Salpeter-IMF-like CMF at the high-mass end. The turnover masses of the CMF depend on the properties of the fibers on which cores/stars begin to form. We suggest that these fibers (\(\mathcal{T}_1\)) are the basic building blocks of star formation, sharing similar properties across different clouds, motivated by the universality of the so-called star formation law. These fibers set a common threshold for star formation, leading to a universal CMF.

\begin{acknowledgements}
X.L. acknowledges the support of the Strategic Priority Research Program of the Chinese Academy of Sciences  under Grant No. XDB0800303,
the National Key R\&D Program of China under Grant No. 2022YFA1603100,
and the National Natural Science Foundation of China (NSFC) through grants No. 12203086. We also thank the anonymous referee for the insightful comments, which helped refine this work. 
\end{acknowledgements}

%

\vspace{5mm}




\bibliography{ms}{}
\bibliographystyle{raa}

\appendix
\section{Structure density profile}
Denote 
$r$ the radius (or width/thickness),
and $\rho$ the density of the $n$-D structures. 
For isothermal cases, the dynamic equilibrium between gravity and the pressure gradient requires that
\begin{equation}
r^{-(2-n)}\int_0^r l^{2-n} \rho \, dl = -A \frac{d \rho}{\rho \, dr}, \label{eq_balance}
\end{equation}
where \( A = c_n\frac{kT}{\mu m_H G} \), \( k \) is the Boltzmann constant, \( G \) is the gravitational constant, 
$\mu$ is the mean molecular weight, $m_H$ is the
mass of hydrogen atom, and \( T \) is the temperature.
Here, $c_n$ is 1, 2, and $2\pi$ for $n=0,1,2$, respectively.
Denote $\rho = e^y$. By differentiating both sides, the above equation (Eq. \ref{eq_balance}) becomes: 
\begin{align} 
\frac{d^2 y}{dr^2} + (2 - n) \frac{dy}{rdr} = -\frac{e^y}{A}. \label{eq_profi} 
\end{align}
For small $r$ close to the center of structure, through expanding \( y \) as 
\begin{equation}
    y = y_0 - \frac{1}{a^2} r^2 + O(r^2)
\end{equation}  and substituting it into Eq.~\ref{eq_profi}, we obtain
\begin{equation}
    (6 - 2n)/a^2 \sim e^{y_0}/A = \rho_0/A. \label{eq_fw}
\end{equation}
Here, \( \rho_0 \) represents the density at \( r = 0 \).
The typical scale of the structures can be estimated
to be $a$ (half width at 1/e of the maximum). From Eq.~\ref{eq_fw}, we have
\begin{equation}
    a \propto A^{1/2}\rho_0^{-1/2}.\label{eq_comwidth}
\end{equation}
For a 2-D sheet (\(n = 2\)), the analytical solution to Eq.~\ref{eq_profi} can be expressed as
\begin{equation}
y(r) = \rho_0 \left( 1 - \tanh^2 \left( \frac{1}{2} \sqrt{\frac{2 \rho_0}{A}} \, r \right) \right). \label{eq_2dsol}
\end{equation}
From Eq.~\ref{eq_rhon}, we have \( \eta_{2d} \propto \rho_0^{1/2} \), which can be directly examined using Eq.~\ref{eq_2dsol}.

\end{CJK*}
\end{document}